\documentclass[%
 reprint,
 amsmath,amssymb,
 aps,
 prl,
 floatfix,
 superscriptaddress,
]{revtex4-2}
\usepackage{graphicx}
\usepackage{dcolumn}
\usepackage{bm}
\usepackage{hyperref}
\hypersetup{hidelinks}
\usepackage{longtable}
\usepackage{booktabs}
\usepackage[table]{xcolor}
\newcommand{\q}[1]{``#1''} 
\newcommand{\SM}{Supplemental Material} 
\begin{document}


\title{Evidence for Hierarchical Turbulent Scaling in Nearby Star-Forming Galaxies}

\author{Luna Maria Scipio}
\affiliation{Computational Science Lab, Informatics Institute, University of Amsterdam, Amsterdam, The Netherlands}

\author{Fernando N\'obrega Santos}
\affiliation{Dutch Institute for Emergent Phenomena (DIEP), Institute for Advanced Studies (IAS), University of Amsterdam, Amsterdam, The Netherlands}
\affiliation{Korteweg--de Vries Institute for Mathematics, University of Amsterdam, Amsterdam, The Netherlands}

\author{Jay Armas}
\affiliation{Institute for Theoretical Physics, University of Amsterdam, Amsterdam, The Netherlands}
\affiliation{Dutch Institute for Emergent Phenomena (DIEP), Institute for Advanced Studies (IAS), University of Amsterdam, Amsterdam, The Netherlands}
\affiliation{Niels Bohr International Academy, The Niels Bohr Institute,
University of Copenhagen, Blegdamsvej 17, DK-2100 Copenhagen \O, Denmark}

\date{\today}

\begin{abstract}
We find recurring signatures of intermittency in molecular-gas velocities from cloud to galactic scales. In 31 selected galaxies from PHANGS-ALMA, velocity-centroid exponents through sixth order show nonlinear order dependence and nearly linear relations between consecutive exponents across galaxies. Positive estimated offsets from single-exponent scaling occur in all 52 galaxies with a valid power law fit and 99 of 100 clouds. A generalized She-Leveque hierarchy accommodates these signatures, supporting a hierarchical description of turbulent scaling in molecular gas across these environments.
\end{abstract}

\maketitle

\section{\label{sec:introduction}Introduction}
Turbulence regulates how gas in the interstellar medium (ISM) is structured, compressed and transported across scales \cite{falceta-goncalvesTurbulenceInterstellarMedium2014}, with consequences for star formation \cite{elmegreenInterstellarTurbulenceObservations2004, maclowControlStarFormation2004}. 
\begin{figure*}[t]
\centering
\includegraphics[width=1\textwidth]{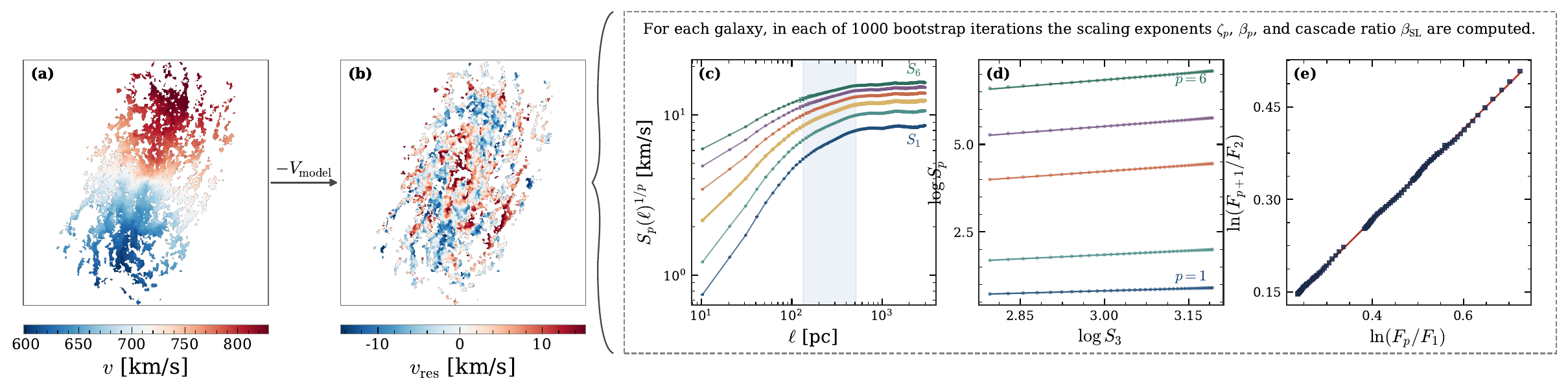}
\caption{Our analysis of each galaxy, illustrated for NGC 3621. Panel (a) shows the line of sight intensity-weighted mean velocity, the \q{moment 1} map from the PHANGS-ALMA dataset. Panel (b) shows the residual map after subtraction of the rotational model $V_{model}$, which removes large scale galaxy kinematics. Panel (c) shows the structure functions $S_p(\ell)^{1/p}$ in $\mathrm{km\,s^{-1}}$ for $p=1$ to $6$. The shaded area marks the fitted range, from twice the beam FWHM to the scale where power law behavior breaks down. Panel (d) shows the extended self-similarity (ESS) exponents $\beta_p$, computed by fitting $\log S_p$ against $\log S_3$. Panel (e) shows the $\beta$ test from \citet{padoanStructureFunctionScaling2003}, whose slope gives the cascade ratio $\beta_\mathrm{SL}$. We repeat the steps in panels (c) to (e) in each of $N=1000$ bootstrap iterations to estimate $\zeta_p$, $\beta_p$ and $\beta_\mathrm{SL}$ with their uncertainties as the 16th and 84th percentiles of the bootstrap distribution.} 
\label{fig:summary}
\end{figure*}
Still, it is not well understood which physical processes drive turbulent motions within the ISM  \cite{krumholzTurbulenceInterstellarMedium2016}. Stellar feedback, on scales of $\sim10$-$100$ pc, is commonly thought to be a major source of turbulent energy injection \cite{elmegreenInterstellarTurbulenceObservations2004, ferrierePlasmaTurbulenceInterstellar2020}. However, another much less investigated driver of turbulence is on the scale of global galactic motions, such as galactic rotation, spiral arm and bar shocks, shear-driven instabilities, and gravitational instabilities \cite{fleckGenerationMaintenanceTurbulence1981, elmegreenInterstellarTurbulenceObservations2004, elmegreen-2024}. Such processes can supply energy at scales of $\sim1$ kpc, where large-scale galactic rotation dominates \cite{ferrierePlasmaTurbulenceInterstellar2020}, with possible transfer down to the molecular cloud scale $< 100$ pc \cite{klessenAccretiondrivenTurbulenceUniversal2010, colmanSignatureLargescaleTurbulence2022}. 

Here we use PHANGS-ALMA velocity-centroid maps to look for signatures of turbulence in molecular gas across nearby galaxies. Across 31 selected galaxies, approximately linear relations between consecutive scaling exponents coexist with nonlinear dependence on order within each galaxy. Together, these patterns support a generalized hierarchical description of turbulent scaling despite differences between galaxies. 

Velocity structure functions (VSFs) characterize the movement of turbulent gas across scales. The VSF of order $p$, defined as
\begin{equation}
S_p(\ell) = \langle |\delta v(\ell)|^p \rangle \propto \ell^{\zeta_p}~~,
\label{eqn:vsf}
\end{equation}
quantifies how velocity fluctuations scale with spatial separation $\ell$. Kolmogorov's foundational K41 theory \cite{kolmogorovLocalStructureTurbulence1991, kolmogorovDissipationEnergyLocally1991, kolmogorov1941degeneration} predicts that in the inertial range of fully developed, isotropic, incompressible turbulence, the exponents $\zeta_p$ scale as $\zeta_p = \frac{p}{3}$. However, real turbulence often deviates from this prediction due to intermittency. Higher-order structure functions give greater weight to intense velocity differences. The nonlinear dependence of their scaling exponents on order characterizes intermittency, showing that intense fluctuations scale differently from more typical motions. To account for this deviation, \citet{sheUniversalScalingLaws1994} proposed a hierarchical model of the turbulent cascade, based on log-Poisson statistics. Here, the statistics of intense velocity fluctuations at each scale are linked to those at the next scale by the cascade ratio $\beta_{\mathrm{SL}}$, sometimes referred to as the intermittency parameter \cite{panPassiveScalarStructures2011}. With the assumption that the third order scaling exponent $\zeta_3$ is unity, the exponents of order $p$ are predicted as
\begin{equation}
    \zeta_p = \frac{p}{9} + C\left(1 - {\beta_{\mathrm{SL}}}^p\right)~~.
    \label{eqn:sl_zeta}
\end{equation}
The cascade ratio $\beta_{\mathrm{SL}}$ is directly related to the co-dimension $C$, where
\begin{equation}
    C = \frac{2/3}{1 - {\beta_{\mathrm{SL}}}^3}~~, \qquad D = 3 - C~~,
    \label{eqn:D_from_beta}
\end{equation}
and $D$ is the Hausdorff dimension, which characterizes the fractal geometry of the structure in which energy dissipation is most intense. For incompressible turbulence, these most dissipative structures are vortex filaments, i.e. $D=1$ and $\beta_{\mathrm{SL}} = (2/3)^{1/3} \approx 0.87$. 
\citet{boldyrevKolmogorovBurgersModelStarforming2002} extended this framework to supersonic turbulence, prevalent in the ISM, with the Kolmogorov-Burgers model. The most intense dissipation in supersonic flows occurs in two-dimensional shock sheets, instead of one-dimensional vortex filaments, i.e. $D = 2$ and $\beta_{\mathrm{SL}} = (1/3)^{1/3} \approx 0.69$.

The She and Leveque hierarchy relates scaling exponents within each turbulent medium. By comparing these exponents across galaxies, we ask which relations persist as the environment and the individual exponents change. Many studies have used VSFs to examine turbulence in individual molecular clouds or cloud samples, often at parsec scales
\cite{ossenkopfTurbulentVelocityStructure2002, boldyrevScalingRelationsSupersonic2002, padoanStructureFunctionScaling2003, chiraHowVelocityStructure2019, ma_examining_2025}. Recently, \citet{ma_examining_2025} examined 167 molecular clouds at scales of  $\sim$0.1 pc to $\sim$20 pc, calculating VSFs up to the third order. They report power law scaling for 60\% of the sample, with relative scaling exponents mostly consistent with the Kolmogorov-Burgers model. At larger circumgalactic scales, \citet{Chen_2024} measured VSFs up to sixth order in eight nebulae around seven QSOs. They compared the relative exponents with K41, She and Leveque, and numerical turbulence predictions. Seven nebulae broadly resemble subsonic predictions, while HE0238$-$1904 is more consistent with the supersonic magnetohydrodynamic prediction of \citet{boldyrevKolmogorovBurgersModelStarforming2002}. 

Still, observational VSF studies remain scarce at the scales between those covered by these cloud and QSO studies. In the Milky Way, \citet{maLocalArmCascade2025} examined Local Arm velocity structure functions and spatial power spectra, with the latter probing approximately $1$-$200$ pc. \citet{henshawUbiquitousVelocityFluctuations2020} investigated molecular-gas VSFs in the nearby galaxy NGC 4321 at these intermediate scales. To our knowledge, no previous study has measured VSFs through sixth order over $\sim$100 pc to 3 kpc across a diverse sample of nearby star forming galaxies. The PHANGS-ALMA survey \cite{leroyPHANGSALMAData2021, leroyPHANGSALMAArcsecond2021}, already used in several studies of molecular gas turbulence \cite{sun-2020, elmegreen-2024, meidt-2025}, makes it possible to address this gap, providing line of sight intensity weighted mean velocity maps of CO(2-1) gas for 90 nearby star forming galaxies at an average resolution of $\sim 100$ pc. 

\begin{figure*}[t]
\centering
\includegraphics[width=1\textwidth]{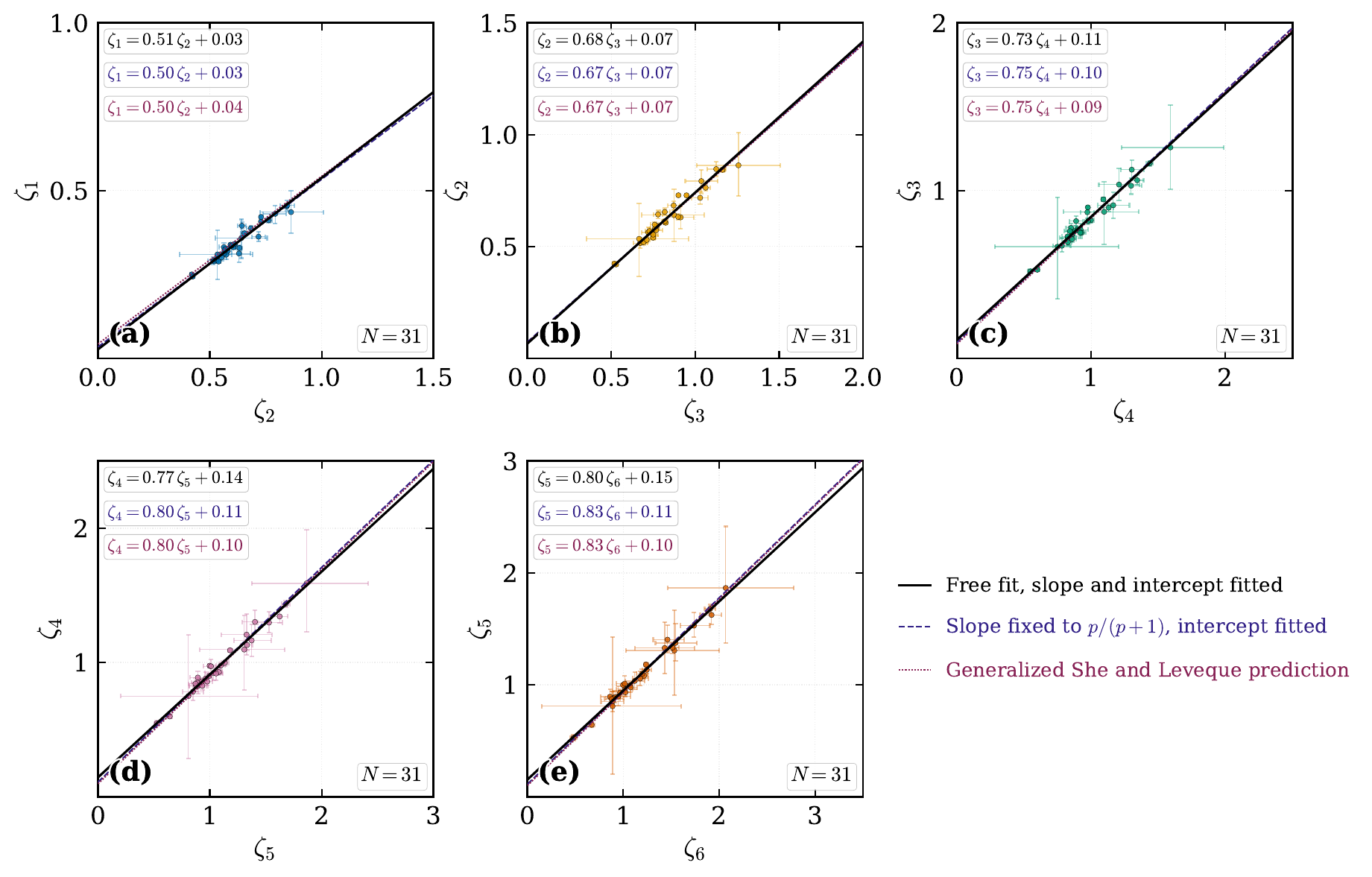}
\caption{Relations between consecutive scaling exponents across the 31 selected PHANGS-ALMA galaxies. Each point represents a pair of exponents measured within one galaxy, with bootstrap error bars. The black line shows our fit of $\zeta_p = m \cdot \zeta_{p+1} + b$, with both slope and intercept free. For the blue line, we fix the slope at $p/(p+1)$ and fit only the intercept in $\zeta_p = \frac{p}{p+1} \zeta_{p+1} + \gamma$. For comparison, the red line evaluates the model at the population medians of $\beta_{SL}$ and $\alpha$.}
\label{fig:interdep}
\end{figure*}

In this letter we analyze VSFs in 67 PHANGS-ALMA galaxies at scales between $\sim$100 pc and 3 kpc. We examine scaling exponents through sixth order and compare them with a generalized She and Leveque model in which both $\zeta_3$ and the singularity exponent $\alpha$ can vary \cite{panPassiveScalarStructures2011}. Of the 67 galaxies, 31 meet the fitting and data quality criteria specified below. We also measure extended self-similarity (ESS), which describes the scaling of each structure function $S_p$ with the third-order function $S_3$, rather than with separation $\ell$ \cite{benziExtendedSelfsimilarityTurbulent1993}. Their relative exponents span the curves found in the transonic to highly supersonic simulations of \citet{panPassiveScalarStructures2011}. As in the molecular clouds studied by \citet{ma_examining_2025}, the measured third-order exponent varies between galaxies. We find approximately linear relations between consecutive scaling exponents across the 31 selected galaxies (\autoref{fig:interdep}) and the 100 molecular clouds (see \SM). In the galaxies, these population relations accompany nonlinear relative-exponent curves (\autoref{fig:betapmedian}). We interpret these complementary observations using the generalized She and Leveque model.

\section{Methods}
We use the strict intensity weighted mean line of sight velocity \q{moment 1} maps and the corresponding error maps of the $12\,\mathrm{m}+7\,\mathrm{m}+\mathrm{TP}$ galaxies from the PHANGS-ALMA pipeline \cite{leroyPHANGSALMAData2021} for the 67 galaxies for which \citet{langPHANGSCOKinematics2020} provide kinematic models. For each galaxy, we remove large scale kinematics by de-projecting the maps and subtracting the rotation model $V_{\text{model}} = V_{\text{sys}} + V_{\text{rot}}(R) \cos\theta \sin i$, with $V_{\text{rot}}(R)$ obtained by linear interpolation of the \citet{langPHANGSCOKinematics2020} rotation curves. As in \citet{Chen_2024}, we then exclude pixels whose velocity uncertainty exceeds three times the median uncertainty, and the $2\%$ most extreme residual velocities on either side of the distribution. From the residual map, we compute velocity increments $\delta v(\ell) = v(\mathbf{r} + \boldsymbol{\ell}) - v(\mathbf{r})$ at each separation $\ell$. We then calculate structure functions $S_p(\ell)$ for orders $p = 1$ to $p = 6$ using \autoref{eqn:vsf}. We correct $S_2$ for noise bias and choose a fitted range that excludes scales dominated by the beam or where power law behavior breaks down (Figure \ref{fig:summary}). Because the inertial scaling range in observational data is often short or noisy, we measure both the raw exponents $\zeta_p$ and the ESS exponents $\beta_p\approx\zeta_p/\zeta_3$. For the latter, we fit log $S_p$ against log $S_3$ \cite{benziExtendedSelfsimilarityTurbulent1993}. The scaling exponents $\zeta_p$ of consecutive orders in \autoref{eqn:sl_zeta} are related through only the cascade ratio $\beta_\mathrm{SL}$ and the order $p$. This framework relies on the constraint of $\zeta_3 = 1$. We find that the galaxies in the PHANGS-ALMA population, as well as the molecular clouds from \citet{ma_examining_2025}, generally yield $\zeta_3 \neq 1$, requiring a broader framework than that of She and Leveque. Therefore, we use the generalized hierarchical form of \citet{panPassiveScalarStructures2011}, written here in terms of $\zeta_3$ and $\alpha$ and derived from the dissipation assumptions stated in the \SM,
\begin{equation}
\zeta_p = \frac{p}{3}\left(\zeta_3 + \alpha\right) + C\left(1 - \beta_{\mathrm{SL}}^{p}\right)~~.
\label{eqn:general_zeta}
\end{equation}
Here $C=-\alpha/(1-\beta_{\rm SL}^3)$ while the earlier expression in \autoref{eqn:D_from_beta} applies to the standard $\alpha=-2/3$ case. Under ideal scaling, defining $\beta_p=\zeta_p/\zeta_3$ and $C'=C/\zeta_3$, Equation \eqref{eqn:general_zeta} gives
\begin{equation}
\beta_p=\frac{p}{3}(1+\frac{\alpha}{\zeta_3})+C'(1-\beta_{\rm SL}^p)~~.
\label{eqn:general_beta}
\end{equation}

By eliminating the term linear in order, we obtain an identity for each galaxy $(i)$
\begin{equation}
\zeta_p^{(i)} = \frac{p}{p+1}\, \zeta_{p+1}^{(i)} + \gamma_p^{(i)}~~,
\label{eqn:interdep}
\end{equation}
with intercept 
\begin{equation}
    \gamma_p^{(i)} = \frac{-\alpha_i}{1-\beta_{SL,i}^3}[(1 - \beta_{SL,i}^p) - \frac{p}{p+1}(1 - \beta_{SL,i}^{p+1})]~~.
    \label{eqn:intercept}
\end{equation}
For a fixed order $p$, each galaxy contributes one point to a plot of $\zeta_p^{(i)}$ against $\zeta_{p+1}^{(i)}$. Equation~\eqref{eqn:interdep} separates its vertical position into a term proportional to $\zeta_{p+1}^{(i)}$ and an offset $\gamma_p^{(i)}$, set by that galaxy's $\alpha_i$ and $\beta_{SL,i}$. If all galaxies have the same offset, their points lie on a line of slope $p/(p+1)$. Different offsets can introduce scatter and, if they vary systematically with $\zeta_{p+1}^{(i)}$, change the population slope. The observed residual $\Delta_{p,i}=\zeta_{p}^{(i)}-[p/(p+1)]\zeta_{p+1}^{(i)}$ removes any contribution proportional to order $p$. It vanishes for single-exponent self-similarity and, within the generalized hierarchy, corresponds to the model offset $\gamma_p^{(i)}$. Thus we fit the population slope and compare it with the reference values $1/2$, $2/3$, $3/4$, $4/5$ and $5/6$ for $p=1,\ldots,5$. The derivation is given in the \SM.

For each galaxy, we estimate the scaling exponents and their uncertainties with a spatial bootstrap. The map is divided into beam sized blocks and one pixel per block is resampled over $N = 1000$ iterations. In each iteration, the first to sixth order exponents $\zeta_p^{(i)}$ and $\beta_p^{(i)}$ are determined, together with the cascade ratio $\beta_{SL,i}$, using the $\beta$-test described in \citet{padoanStructureFunctionScaling2003}. These steps are illustrated for NGC 3621 in \autoref{fig:summary}. Keeping the cascade ratio $\beta_{SL}^{(i)}$ fixed, we fit \autoref{eqn:general_beta} for $\alpha_{i}$ and compute the model-inferred dimension as $D=3 + \alpha_{i}/(1-\beta_{SL,i}^3)$. Each galaxy's measured ESS exponents are compared with the predictions of \citet{panPassiveScalarStructures2011} by summing the squared differences between the observed and simulated $\beta_p^{(i)}$ values, each divided by the squared uncertainty of the observed value. This weighted $\chi^2$ ranks how closely each simulation curve matches the galaxy's measured exponents. Each galaxy is labeled by the Mach number of the matching simulation, or by a range when its exponents lie between simulation curves (see \autoref{fig:mach_classification} and the detailed rule in the \SM). Additionally, we fit the population relation for each pair of consecutive orders using a maximum likelihood method based on distances perpendicular to the fitted line. This method accounts for each galaxy's bootstrap covariance. We first fit both the slope and intercept, then fix the slope at $p/(p+1)$ and fit only the intercept (see \SM). 

The quality of the \q{moment 1} maps and of the kinematic subtractions varies widely among the 67 galaxies. Some galaxies have few usable velocity measurements, while others retain substantial contributions from non-turbulent motions. To compare the exponents with the generalized She and Leveque model in \autoref{eqn:general_zeta}, we select galaxies based on the quality of their kinematic models. To limit residual noncircular motions and projection effects, we select galaxies whose kinematic orientation \citet{langPHANGSCOKinematics2020} classify as reliable, for which they provide a smooth fit to the rotation curve, and whose inclination satisfies $i + \sigma_i \leq 70^\circ$. Of the 67 galaxies, 52 yield a valid fit range of at least five lags over which power law scaling occurs. Of these, 31 meet the quality criteria. Their exponents are listed in \autoref{tab:scaling-exponents}. 
\section{Results}

Our fitted intervals lie between 58 pc and 2982 pc, with median lower and upper bounds of $234$ pc and $590$ pc. The median ESS curve follows the Kolmogorov-Burgers prediction up to fourth order and lies slightly below it at higher orders: $\beta_2\simeq0.748$ and $\beta_6\simeq1.446$, compared with $0.741$ and $1.556$ for Kolmogorov-Burgers and $2/3$ and $2$ for single-exponent self-similarity (\autoref{fig:betapmedian}). We find a range of individual exponent curves, with cascade ratios from $0.61$ to $0.78$. Comparison with the numerical curves of \citet{panPassiveScalarStructures2011} gives estimated Mach numbers from $\mathcal{M}_s\simeq0.9$ to $\mathcal{M}_s\simeq6.1$, spanning transonic to highly supersonic regimes (\autoref{fig:mach_classification} and \autoref{tab:scaling-exponents}). This spread motivates the generalized hierarchy and the population comparison in \autoref{fig:interdep}. We obtain slopes of 0.51, 0.68, 0.73, 0.77 and 0.80, close to the reference values $p/(p+1)$. We show the line evaluated at the sample medians $\beta_{SL}=0.73^{+0.03}_{-0.06}$ and $\alpha=-0.71^{+0.27}_{-0.18}$ in red in \autoref{fig:interdep}. The quoted intervals describe the 16th-84th percentile spread across galaxies. Our central estimates of the dimension of the most intense dissipative structures range from $D=1.26$ in NGC 4321 and NGC 3627 to $D=3.04$ in NGC 3507, with population median $D=1.93^{+0.35}_{-0.40}$ (\autoref{tab:scaling-exponents}). These intervals again describe the spread across the 31 selected galaxies. Noninteger values describe fractal dissipative structures. The median Hausdorff dimension $D$ near two is compatible with sheet-like concentrations of dissipation, strengthening the Kolmogorov-Burgers comparison for the median. Beyond the slopes, the raw-exponent residual $\Delta_2=\zeta_2-2\zeta_3/3$ is positive in all 52 galaxies with accepted fits, including the 21 outside the final subset. This positive offset therefore also occurs outside the final 31 galaxy subset. The calculation details are described in the \SM.

\section{Discussion} 
The population relations in \autoref{fig:interdep} and the nonlinear relative-exponent curves in \autoref{fig:betapmedian} together support a hierarchical description of galactic velocity-centroid scaling. The single-exponent family $\zeta_{p}^{(i)}=pH_i$ provides a useful reference, as it gives population slopes $p/(p+1)$, zero offsets and relative exponents $\beta_p=p/3$, even when $H_i$ varies between galaxies. The Kolmogorov K41 theory is its $H=1/3$ case. Our positive offsets and nonlinear exponent curves connect the population relations to intermittency. Positive offsets in the broader 52 galaxy sample extend this evidence beyond the subset selected by kinematic and inclination criteria. Within the generalized hierarchy, differences between exponent sequences can be interpreted through parameters describing intermittency and the concentration of intense dissipation.

Evidence for turbulent cascades is also present in the molecular cloud sample of \citet{ma_examining_2025}, providing an independent observational comparison of low-order population scaling at parsec scales. Our fits to consecutive exponents of the 100 molecular clouds studied by \citet{ma_examining_2025} give slopes near $p/(p+1)$, while 99 of the 100 clouds have positive offsets $\Delta_2$ and their median raw ratio is $\zeta_2/\zeta_3\simeq0.749$, which is the counterpart of the ESS exponent $\beta_2$. The shared positive offsets identify a recurring departure from single-exponent self-similarity in these molecular-gas samples. 

Connecting projected centroids to three-dimensional turbulence requires a treatment of density weighting and projection \cite{esquivelStatisticsVelocityCentroids2007}. To establish a cascade mechanism, we would also need velocity statistics sensitive to energy transfer and independent constraints on the driving and dissipative processes. To quantify the evidence for the generalized hierarchy, we need joint exponent uncertainties and comparisons with alternative models passed through the same projection, beam, noise, fitting and selection procedures. These checks would show how residual noncircular motions, limited fitted ranges and rare velocity differences affect our results. They would also account for the use of sample-derived parameters in the red reference line. K41-type scaling and the Gaussian example in the \SM\ illustrate why the population slopes and pooled $\beta$-test must be assessed together with the exponent curves and offsets.

Interdependent exponents also occur in the human brain \cite{castroInterdependentScalingExponents2025} and urban populations \cite{merbisHyperscalingSpatialFluctuations2026}, illustrating the broader organization of scaling in complex systems \cite{lopesPowerLawBehaviour2018}. These mathematical parallels encourage us to compare relations among exponents in different systems. In the cases studied here, the physical interpretation rests on velocity statistics and their departures from single-exponent self-similarity.

We find interdependent velocity-scaling exponents and nonlinear order dependence in the selected PHANGS-ALMA galaxies. Our results are compatible with a generalized hierarchy of turbulent intermittency. Positive low-order offsets in the broader galaxy sample and in molecular clouds extend these signatures across scales. Our findings invite us to test which relations among exponents persist as astrophysical conditions change, and which depend on the turbulent regime. By comparing diverse galaxies, we can investigate what is universal in turbulent scaling.

\begin{acknowledgments}
\textit{Acknowledgements--} F. A. Nobrega Santos is supported by the Dutch Institute for Emergent Phenomena (DIEP) cluster at the University of Amsterdam under the Research Priority Area Emergent Phenomena in Society: Polarization, Segregation and Inequality. J. Armas is partly funded by the DIEP cluster at the University of Amsterdam via the DIEP programme Foundations and Applications of Emergence (FAEME) and the national NWA- ORC 2023 consortium EAAS.

We use data processed by the PHANGS-ALMA survey \cite{leroyPHANGSALMAData2021, leroyPHANGSALMAArcsecond2021}, which relies on the following ALMA data: \nolinkurl{ADS/JAO.ALMA/\#2012.1.00650.S}, \nolinkurl{ADS/JAO.ALMA/\#2015.1.00925.S}, \nolinkurl{ADS/JAO.ALMA/\#2015.1.00956.S}, \nolinkurl{ADS/JAO.ALMA/\#2017.1.00886.L},
\nolinkurl{ADS/JAO.ALMA/\#2018.1.01651.S}.
ALMA is a partnership of ESO (representing its member states), NSF (USA) and NINS
(Japan), together with NRC (Canada), MOST and ASIAA (Taiwan), and KASI (Republic
of Korea), in cooperation with the Republic of Chile. The Joint ALMA Observatory is
operated by ESO, AUI/NRAO and NAOJ. The National Radio Astronomy Observatory is
a facility of the National Science Foundation operated under cooperative agreement by
Associated Universities, Inc.

The authors used OpenAI Codex Astra 6 Pro and Claude Code Opus 5.5 to assist with manuscript revision, literature cross-checks, and review of mathematical arguments and analysis code. The authors performed human checks on the AI-assisted material and take responsibility for the final content and conclusions.
\end{acknowledgments}

\bibliography{references}
\onecolumngrid 
\clearpage
\section{Supplemental Material}
\setcounter{equation}{0}
\renewcommand{\theequation}{S\arabic{equation}}
\renewcommand{\theHequation}{S\arabic{equation}}
\setcounter{figure}{0}
\renewcommand{\thefigure}{S\arabic{figure}}
\renewcommand{\theHfigure}{S\arabic{figure}}
\setcounter{table}{0}
\renewcommand{\thetable}{S\arabic{table}}
\renewcommand{\theHtable}{S\arabic{table}}

\subsection{Generalization of She and Leveque}
Here we derive the generalized scaling relation used in the letter, and also considered in \cite{panPassiveScalarStructures2011}. The She and Leveque model \cite{sheUniversalScalingLaws1994} describes how the scaling of increasingly high velocity moments reflects increasingly intense dissipation fluctuations. Its original prediction is
\begin{equation}
\zeta_p=\frac{p}{9}+C(1-\beta_{\mathrm{SL}}^p)~~,
\qquad C=2,\quad \beta_{\mathrm{SL}}=(2/3)^{1/3}~~.
\label{eqn:SSL}
\end{equation}
The term proportional to $p$ cancels when we subtract $p\zeta_{p+1}/(p+1)$ from $\zeta_p$. Thus
\begin{equation}
\zeta_p=\frac{p}{p+1}\zeta_{p+1}+\gamma_p~~,
\qquad
\gamma_p=C\left[(1-\beta_{\mathrm{SL}}^p)-\frac{p}{p+1}(1-\beta_{\mathrm{SL}}^{p+1})\right]~~.
\label{eqn:SMinterdep}
\end{equation}
The same cancellation will hold for the generalized model below, because its first term is also linear in $p$.

To allow the measured third-order exponent to differ from unity, we follow the generalized refined similarity relation of \citet{BENZI1996162}. Let $\epsilon_\ell$ be the dissipation rate per unit mass averaged over a region of size $\ell$. Its moments describe how strongly dissipation fluctuates between regions at that scale. The relation connects these dissipation moments to velocity structure functions:
\begin{equation}\label{eqn:GESS}
\frac{S_p(\ell)}{S_3(\ell)^{p/3}}
\sim\frac{\langle\epsilon_\ell^{p/3}\rangle}{\langle\epsilon_\ell\rangle^{p/3}}~~.
\end{equation}
Both sides compare the moment of interest with a reference moment. In particular, $\sim$ denotes proportionality with a factor independent of $\ell$ in the fitted range. Using the measured $S_3(\ell)$ as the reference leaves its exponent $\zeta_3$ free. We adopt this relation as a model connecting the centroid structure functions to dissipation fluctuations. This is the physical step beyond extended self-similarity (ESS) \cite{benziExtendedSelfsimilarityTurbulent1993}: measuring ESS exponents requires only fitting $S_p$ against $S_3$, whereas interpreting them through dissipation requires the relation above. Here all $S_p$, including $S_3$, are absolute moments as defined in \autoref{eqn:vsf}.

We now write $S_p(\ell)\sim\ell^{\zeta_p}$ and $M_q(\ell)\equiv\langle\epsilon_\ell^q\rangle\sim\ell^{\tau_q}$. The left side of \autoref{eqn:GESS} then scales as $\ell^{\zeta_p-p\zeta_3/3}$ and the right side as $\ell^{\tau_{p/3}-p\tau_1/3}$. Equating these powers gives
\begin{equation}\label{eqn:zetaESS}
\zeta_p=\frac{p}{3}\zeta_3+\tau_{p/3}-\frac{p}{3}\tau_1~~.
\end{equation}
It remains to specify the dissipation exponents $\tau_q$. Following She and Leveque, we define the ratio of successive dissipation moments $E_q=M_{q+1}/M_q$. Higher $q$ places more weight on regions with intense dissipation, and $E_\infty$ describes their limiting intensity. The She and Leveque hierarchy assumes
\begin{equation}\label{eqn:smSLsym}
E_{q+1}(\ell)=A_q E_q(\ell)^b E_\infty(\ell)^{1-b}~~,
\qquad E_\infty(\ell)\sim\ell^\alpha,\qquad b=\beta_{\mathrm{SL}}^3~~,
\end{equation}
where $A_q$ is independent of scale and $0<b<1$. The exponent $\alpha$ describes how the strongest dissipation fluctuations change with scale. The dissipation hierarchy uses $b$, while the velocity formula uses $\beta_{\mathrm{SL}}$. Their cubic relation arises because velocity order $p$ is paired with dissipation order $q=p/3$ in \autoref{eqn:zetaESS}.

Since $E_q\sim\ell^{\tau_{q+1}-\tau_q}$, matching powers of $\ell$ in \autoref{eqn:smSLsym} gives
\begin{equation}
\tau_{q+2}-\tau_{q+1}=b(\tau_{q+1}-\tau_q)+(1-b)\alpha~~.
\end{equation}
Defining $d_q=\tau_{q+1}-\tau_q$, this recurrence becomes $d_{q+1}-\alpha=b(d_q-\alpha)$: each successive difference approaches $\alpha$ by a factor $b$. The normalization $M_0=1$ gives $\tau_0=0$. We also assume that the mean dissipation $M_1$ is independent of the averaging scale, giving $\tau_1=0$ and hence $d_0=0$. Repeated substitution yields $d_q=\alpha(1-b^q)$ while summing these differences from zero to $q-1$ gives, for integer $q\geq1$,
\begin{equation}
\tau_q=\sum_{j=0}^{q-1}d_j
=\alpha\left[q-\frac{1-b^q}{1-b}\right]
=\alpha q+C(1-b^q),\qquad C=-\frac{\alpha}{1-b}~~.
\end{equation}
We use the same expression for real as well as integer moment orders, following the log-Poisson model, as required to evaluate $\tau_{p/3}$ \cite{sheUniversalScalingLaws1994,BENZI1996162}. Inserting it into \autoref{eqn:zetaESS}, with $\tau_1=0$ and $b=\beta_{\mathrm{SL}}^3$, gives
\begin{equation}
\zeta_p=\frac{p}{3}(\zeta_3+\alpha)+C(1-\beta_{\mathrm{SL}}^p)~~,
\qquad C=-\frac{\alpha}{1-\beta_{\mathrm{SL}}^3}~~.
\label{eqn:SM13}
\end{equation}
At $p=3$, the terms $\alpha$ and $C(1-\beta_{\mathrm{SL}}^3)$ cancel, so the expression returns the specified $\zeta_3$. At $\zeta_3=1$, $\alpha=-2/3$ and $b=2/3$, it recovers the original She and Leveque result in \autoref{eqn:SSL}. The generalized expression retains the same hierarchy while allowing the measured third-order scaling and the exponent of the strongest dissipation fluctuations to vary. A related formulation is used by \citet{Chen_2024}.

Kolmogorov's K41 scaling is the nonintermittent special case of this generalized She and Leveque framework. Setting $\alpha=0$ gives $C=0$ for $0<\beta_{\mathrm{SL}}<1$, so \autoref{eqn:SM13} reduces to $\zeta_p=(p/3)\zeta_3$. With $\zeta_3=1$, this is precisely the K41 prediction $\zeta_p=p/3$. Consequently,
\[
\zeta_p-\frac{p}{p+1}\zeta_{p+1}
=\frac{p}{3}-\frac{p}{p+1}\frac{p+1}{3}=0~~.
\]
Thus K41 gives the same coefficient $p/(p+1)$ between consecutive orders, with zero intercept $\gamma_p$. The generalized nonintermittent case, with $\zeta_3=3H$, gives $\zeta_p=pH$ and also has zero intercept. Allowing $H$ to vary between galaxies produces a population line of slope $p/(p+1)$. Strict K41 fixes $H=1/3$ and gives the same exponent pair for every galaxy. K41 is therefore a special case of the generalized framework, rather than of the original fixed-parameter intermittent prediction in \autoref{eqn:SSL}.

For exact power-law scaling and nonzero $\zeta_3$, dividing by $\zeta_3$ gives the relative exponents $\beta_p=\zeta_p/\zeta_3$:
\begin{equation}
\beta_p=\frac{p}{3}\left(1+\frac{\alpha}{\zeta_3}\right)+C'(1-\beta_{\mathrm{SL}}^p)~~,
\qquad C'=C/\zeta_3~~.
\label{eqn:sm13pan}
\end{equation}
In the notation of \citet{panPassiveScalarStructures2011}, the coefficient of $p$ is $\gamma'=(1+\alpha/\zeta_3)/3$ and $C'=(1-3\gamma')/(1-\beta_{\mathrm{SL}}^3)$. Thus their expression has the same form, with $\alpha=\zeta_3(3\gamma'-1)$. Finally, subtracting $p\zeta_{p+1}/(p+1)$ from \autoref{eqn:SM13} cancels the entire term containing $\zeta_3+\alpha$, leaving
\begin{equation}
\gamma_p=\frac{-\alpha}{1-\beta_{\mathrm{SL}}^3}
\left[(1-\beta_{\mathrm{SL}}^p)-\frac{p}{p+1}(1-\beta_{\mathrm{SL}}^{p+1})\right]~~.
\label{eqn:SMintercept}
\end{equation}
This is the offset in \autoref{eqn:SMinterdep}. When it is applied to a population, $\alpha$, $\beta_{\mathrm{SL}}$ and hence $\gamma_p$ take their fitted values for each galaxy, as indicated by the index $i$ in the main text.

\subsection{Supplemental methods}
Here we explain the numerical implementation that leads to the main results in the letter. Since the rotation curves of \citet{langPHANGSCOKinematics2020} are tabulated in kpc, we convert the radius of each pixel to kpc using the same distances as used by \citet{langPHANGSCOKinematics2020}, rather than the more recent distances provided with the PHANGS-ALMA data pipeline. The rotation curve starts at $V_{\rm rot}(0)=0$, is linearly interpolated between the tabulated radii and is held constant beyond the last tabulated point. After the removal of the rotational model, velocity increments are computed from the residual velocity map at each spatial separation $\ell$ as
\begin{equation}
    \delta v(\ell) = v(\mathbf{r} + \boldsymbol{\ell}) - v(\mathbf{r})~~,
    \label{eqn:incr}
\end{equation}
 where $\mathbf{r}$ is the coordinate of each pixel containing a valid velocity measurement. For ease of computation, we express lags as integers in pixel units, starting from $\ell = 1$ pixel and increasing by one pixel at each step. The physical size of one pixel varies per galaxy. Due to the finite number of square pixels, for each lag $\ell$, all pixel pairs separated by a Euclidean distance in the interval $[\ell - \tfrac{1}{2},\, \ell + \tfrac{1}{2})$ pixels are collected across all orientations. The computation stops at the first lag for which the total number of pairs falls below 1000, or when the physical separation exceeds 3 kpc. For each pair, the propagated uncertainty is $\sigma_\mathrm{pair} = \sqrt{\sigma_1^2 + \sigma_2^2}$, where $\sigma_1$ and $\sigma_2$ are the pixel uncertainties from the error maps provided by the PHANGS-ALMA pipeline. The structure functions of orders $p = 1$ to $p = 6$ are computed as
\begin{equation}
    S_p(\ell) = \langle |\delta v(\ell)|^p \rangle~~.
    \label{eqn:SMvsf}
\end{equation}
 After the computation of the structure functions,  a noise bias correction is applied to the second order structure function $S_2$ by subtracting the pair-noise variance
\begin{equation}
    S_2^\mathrm{corr}(\ell) = S_2^\mathrm{obs}(\ell) - \langle \sigma_\mathrm{pair}^2 \rangle~~,
    \label{eqn:SMs2_bias_corr}
\end{equation}
where $\langle \sigma_\mathrm{pair}^2 \rangle = \bigl\langle \sigma_1^2 + \sigma_2^2 \bigr\rangle$ is the sum of the two pixel variances, averaged over all pairs at lag $\ell$ \cite{zuhoneMAPPINGGASTURBULENCE2016, cucchettiMappingTurbulenceIntracluster2019, henshawUbiquitousVelocityFluctuations2020}. This expression assumes independent measurement errors at the two pixels, instead correlated errors would add a covariance term. Only the second order receives this explicit correction. Subsequently, for each galaxy a fit range is determined to exclude scales where beam smearing dominates, with a lower bound $\ell_{\text{min}} = 2 \times \mathrm{FWHM}_{\text{beam}}$, and an upper bound is determined by requiring that $S_1$, $S_2$ and $S_3$ each individually follow a log-log power law with $R^2 \geq 0.99$. A valid fit range must contain at least five lags.  

In observational data the inertial scaling range is often short or noisy and does not show a clean power-law scaling. This can be due to a short fit range, residuals of large-scale kinematics, and effects of the beam. To address this limitation, \citet{benziExtendedSelfsimilarityTurbulent1993} developed the method of extended self-similarity, which fits one structure function against another when scaling against lag is limited. Therefore, both the raw exponents and ESS exponents $\beta_{p}$ are computed. The scaling exponents and corresponding uncertainties are estimated with a spatial bootstrap, similar to the modified bootstrap employed in \citet{chenEmpiricalConstraintsTurbulence2022}. A bootstrap is used, because only taking the propagated $\sigma_\mathrm{pair}$ errors into account is insufficient, because adjacent pixels are spatially correlated, particularly at separations below the beam size. To account for the spatial correlation, the residual velocity map is divided into square blocks with sides equal to the beam FWHM, expressed in integer pixel units. Each grid cell contains at least one valid pixel and is treated as a sampling block. Selecting one pixel per beam-sized block reduces direct oversampling of the beam. In each bootstrap iteration, one pixel is selected uniformly at random from each block, and its velocity is perturbed by adding Gaussian noise drawn from the local measurement uncertainty

\begin{equation}
    v'_i = v_i + \epsilon_i, \qquad
    \epsilon_i \sim \mathcal{N}(0,\,\sigma_{i}^{2})~~,
    \label{eqn:boot_perturb}
\end{equation}

where $\sigma_{i}$ is the velocity uncertainty at pixel $i$, thus the noise on the \q{moment 1} map. The structure functions are then recomputed from all pairs formed between the selected pixels, with the noise bias correction of \autoref{eqn:SMs2_bias_corr} applied to $S_2$ in each iteration. The error estimate used to weight ${S_p}(\ell_i)$ is
\begin{equation}
    \sigma_{S_p}(\ell_i) = \frac{\mathrm{std}(|\delta v(\ell_i)|^p)}{\sqrt{N_\mathrm{pairs}}}~~,
\end{equation}
 following \citet{henshawUbiquitousVelocityFluctuations2020}.
The scaling exponents first to sixth order $\zeta_p$ are obtained within the previously determined valid fit range by weighted least squares in log-log space, and ESS exponents $\beta_p$ by fitting $\log S_p \propto \beta_p \log S_3$ with orthogonal distance regression. Similarly, within each iteration the cascade ratio $\beta_{SL}$ is obtained with the $\beta$-test from \citet{padoanStructureFunctionScaling2003}. This test constructs the intensity ratios in \autoref{eqn:intensityratios},
\begin{equation}
    F_p(\ell) = S_{p+1}(\ell)/S_p(\ell)~~,
    \label{eqn:intensityratios}
\end{equation}
 for $p=1,\ldots,5$ and all lags $\ell$ within the fit range. If the hierarchical model holds and the order-dependent prefactors permit a common line, the points 
 \begin{equation}
     \bigl(\ln(F_p/F_1),\; \ln(F_{p+1}/F_2)\bigr)~~,
     \label{eqn:points}
 \end{equation}
 across all orders and lags fall onto a single straight line whose slope is the cascade ratio $\beta_{\mathrm{SL}}$. We fit this line by ordinary least squares with a free intercept. With the cascade ratio $\beta_{SL}$ fixed, \autoref{eqn:general_beta} is fitted with least squares for $\alpha$. The corresponding co-dimension is then computed as $C=\frac{-\alpha}{1-\beta_{SL}^3}$ and subsequently the Hausdorff dimension is computed as  $D=3-C$. This procedure is repeated $N = 1000$ times. The reported exponents are the median of the bootstrap distribution, and the lower and upper uncertainties are taken as the distances from the median to the 16th and 84th percentiles, respectively. 

The following example explains why the slope and linearity of the pooled $\beta$-test must be considered together with the measured exponent curves. For a fixed-shape increment distribution $\delta v(\ell)\overset{d}{=}a(\ell/\ell_0)^H X$ with finite absolute moments, $S_p(\ell)=\langle|X|^p\rangle a^p(\ell/\ell_0)^{pH}$, hence $\zeta_p=pH$, $\beta_p=p/3$ and $\Delta_p=0$ (for nonzero $H$). If $X$ is standard Gaussian, $\langle|X|^p\rangle=2^{p/2}\Gamma[(p+1)/2]/\sqrt{\pi}$. All lag and amplitude factors cancel from \autoref{eqn:points}, leaving only three order-dependent points for $p=2,3,4$, repeated at each lag. Their free-intercept least-squares fit has slope $0.7746$ and $R^2=0.9997$.  Forcing a zero intercept gives a slope $0.7181$ and centered around $R^2=0.9939$. These exact-moment examples exceed the $0.99$ threshold under both conventions, without testing a finite map or the complete selection pipeline. They do not reproduce the observed order curvature. Thus a high $R^2$ alone cannot distinguish this null from an intermittent field. 
\begin{figure}[h]
\centering
\includegraphics[width=0.5\linewidth]{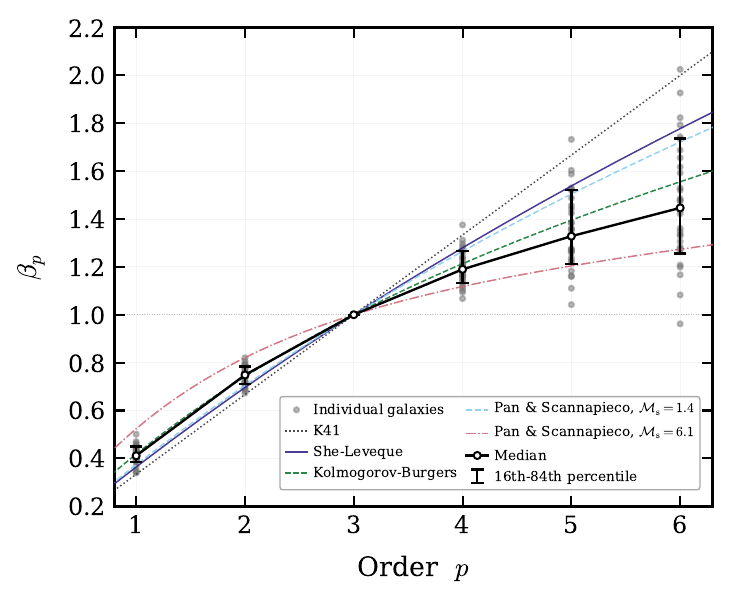}
\caption{The median ESS scaling $\beta_p$ versus order $p$ for all 31 galaxies (black line). The bars show the 16th-84th percentile spread across galaxies. Faint grey points show individual fitted estimates. Theoretical predictions from K41 \cite{kolmogorovLocalStructureTurbulence1991, kolmogorovDissipationEnergyLocally1991, kolmogorov1941degeneration}, \citet{sheUniversalScalingLaws1994} and Kolmogorov-Burgers \cite{boldyrevKolmogorovBurgersModelStarforming2002} are included, along with the numerical simulation results at Mach numbers $\mathcal{M}_s =1.4$ and $\mathcal{M}_s= 6.1 $ from \citet{panPassiveScalarStructures2011}.}
\label{fig:betapmedian}
\end{figure}
\begin{figure}[h!]
    \centering
    \includegraphics[width=0.5\linewidth]{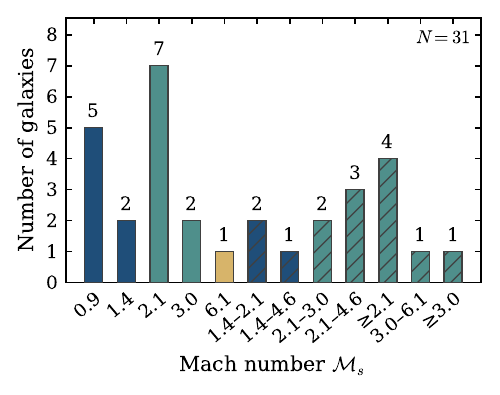}
    \caption{The 31 selected galaxies grouped by their classification against the six \citet{panPassiveScalarStructures2011} Mach number curves, from a weighted $\chi^2$ fit to each galaxy's ESS $\beta_p$ curve. Each label is the Mach number of the assigned simulation. A range is shown when two simulation curves bracket the observed exponents.}
\label{fig:mach_classification}
\end{figure}
\autoref{fig:betapmedian} shows the median scaling of the ESS exponents, along with the theoretical predictions by K41, She and Leveque and Kolmogorov-Burgers. The median values at orders one to six are approximately $(0.41, 0.75, 1, 1.19, 1.33, 1.45)$. All 31 fitted central estimates lie above $p/3$ below third order and below it above third order. We note that $\beta_3=1$ is a normalization identity. In \autoref{tab:scaling-exponents}, 29 of the 31 reported individual 16th-84th percentile intervals for $\beta_2$ lie wholly above $2/3$. These intervals describe the resampling variation for individual galaxies. The bars in \autoref{fig:betapmedian} describe the spread between galaxies. With the $\beta_p$ values obtained for each galaxy, we can compare them to the predictions of \citet{panPassiveScalarStructures2011}, the result of which is visible in \autoref{fig:mach_classification}. This is done for each galaxy by a weighted $\chi^2$ computed against each of the predicted curves of \citet{panPassiveScalarStructures2011}. This $\chi^2$ is weighted by the inverse square of the average of the upper and lower uncertainties of each fitted $\beta_p$. The fixed normalization $\beta_3=1$ supplies no independent constraint. For the best-fitting curve, for which the $\chi^2$ is lowest, the nominal p-value is checked to be above 0.05 as part of the rule for assigning a simulation curve. If it is below this value, it is checked whether the data falls between the two best-fitting curves. If not, this range is widened, curve by curve, until the data is bounded between two curves. When the data exceeds the shallowest or steepest curve, this is left as an open bound. The classification results are included in \autoref{tab:scaling-exponents}. The assigned Mach number identifies the simulation whose exponent curve most closely resembles the observed one. A range identifies the simulation curves that bracket the measurements. The statistical and physical interpretation of these assignments is discussed in the main text.

To fit the relations between consecutive exponents across galaxies, we account for uncertainties in both exponents and their covariance within each galaxy. Instead of fitting the classical $y=mx+b$, we perform a maximum likelihood optimization using orthogonal displacement
\citep{hogg_data_2010}, with $\theta$ and $b^\perp = b \cos \theta$, where $m=\tan\theta$. 
We define $\hat{\mathbf{v}} = (-\sin\theta,\cos\theta)$, orthogonal to the line of the linear relation. For each galaxy $i$ with scaling exponents
$\mathbf{Z}_i = (\zeta_{p+1,i}, \zeta_{p,i})$ and $2\times2$ bootstrap
covariance $S_i$, the orthogonal displacement $\Delta_i$ and its variance $\Sigma^2_i$ are
\begin{equation}
\Delta_i = \hat{\mathbf{v}}^\top\mathbf{Z}_i - b^\perp~~,
\qquad
\Sigma_i^2 = \hat{\mathbf{v}}^\top\mathbf{S}_i\,\hat{\mathbf{v}}~~,
\end{equation}

To account for intrinsic scatter, we introduce $V$. The fit is now given by maximizing
\begin{equation}
\ln \mathcal{L} = K
- \sum_i \frac{1}{2}\ln\!\left(\Sigma_i^2 + V\right)
- \sum_i \frac{\Delta_i^2}{2\left[\Sigma_i^2 + V\right]}~~,
\end{equation}
over $(\theta, b^\perp, V)$, which gives the slope, intercept and intrinsic scatter. We similarly employ this method for the fit of the intercept $\gamma$, with the slope fixed as $p/(p+1)$.

\subsection{Full set of 52 galaxies}
\begin{figure*}[h]
    \centering
    \includegraphics[width=1\linewidth]{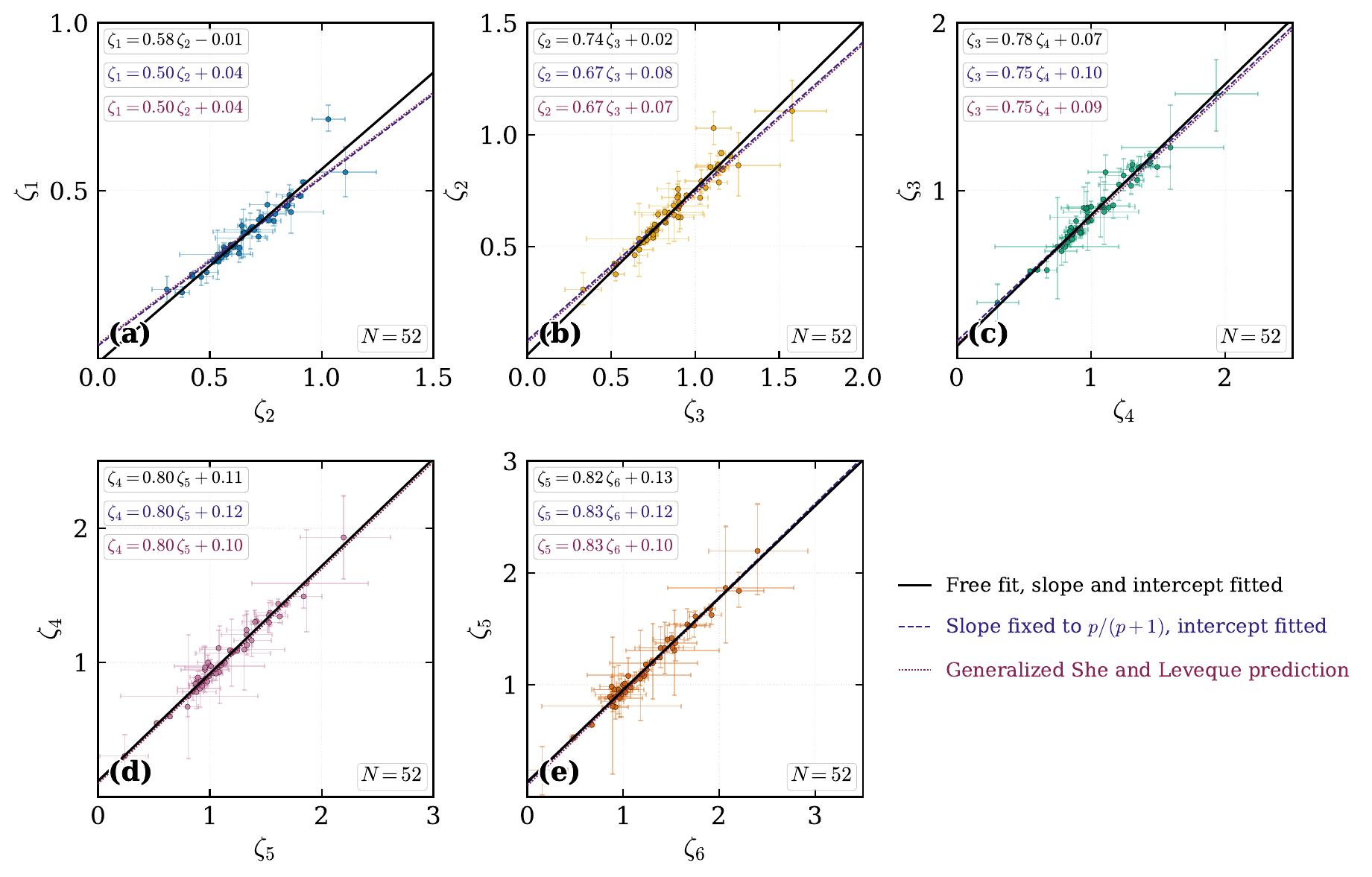}
    \caption{Linear interdependence of consecutive scaling exponents of the 52 PHANGS-ALMA galaxies that yield power law scaling. The panels show $\zeta_p$ versus $\zeta_{p+1}$ for each galaxy, with error bars from the bootstrap procedure. The black line shows our fit of $\zeta_p = m \cdot \zeta_{p+1} + b$, with slope and intercept free. The blue line fixes the slope at $p/(p+1)$ and fits only the intercept in $\zeta_p = \frac{p}{p+1} \zeta_{p+1} + \gamma$. The red line evaluates the model at the medians of $\beta_{SL}$ and $\alpha$ from the 31 selected galaxies.}
    \label{fig:fullsetinterdependence}
\end{figure*}

When removing the selection criteria for the 31 galaxies and fitting the consecutive scaling exponents for the 52 galaxies that yield a power law, a similar linear relation is retrieved, which is visible in \autoref{fig:fullsetinterdependence}. The full-sample fit shows how the population relation changes when the selection based on data quality is relaxed. The directly encoded second-to-third-order pairs give $\Delta_2>0$ in all galaxies (median $\Delta_2\simeq0.077$), including the 21 additional galaxies outside the final subset (median $\Delta_2\simeq0.085$). The offsets are calculated from the fitted central values. Their positive sign therefore also occurs without the selection of galaxies.
\clearpage
\section{Results per galaxy}
\begingroup
\centering
\footnotesize
\setlength{\LTcapwidth}{\textwidth}
\setlength{\tabcolsep}{1.9pt}
\renewcommand{\arraystretch}{1.15}
\setlength{\LTleft}{0pt}
\setlength{\LTright}{0pt}
\begin{longtable*}{@{\extracolsep{\fill}}l c c c c c c c c@{}}
\toprule
Galaxy & $\beta_{\mathrm{SL}}$ & $\alpha$ & $D$ & $\zeta_2$ & $\beta_2$ & $\zeta_3$ & $\mathcal{M}_s$ classification & Fit range [pc] \\
\midrule
\endfirsthead
\toprule
Galaxy & $\beta_{\mathrm{SL}}$ & $\alpha$ & $D$ & $\zeta_2$ & $\beta_2$ & $\zeta_3$ & $\mathcal{M}_s$ classification & Fit range [pc] \\
\midrule
\endhead
\midrule
\endfoot
\midrule
\endlastfoot
IC\,1954 & $0.71^{+0.01}_{-0.01}$ & $-0.77^{+0.07}_{-0.08}$ & $1.79^{+0.13}_{-0.13}$ & $0.58^{+0.02}_{-0.02}$ & $0.76^{+0.01}_{-0.01}$ & $0.76^{+0.03}_{-0.03}$ & $2.1\text{--}4.6$ & $217\text{--}590$ \\
IC\,5273 & $0.70^{+0.02}_{-0.02}$ & $-0.30^{+0.51}_{-0.50}$ & $2.56^{+0.77}_{-0.79}$ & $0.86^{+0.14}_{-0.14}$ & $0.67^{+0.04}_{-0.04}$ & $1.26^{+0.25}_{-0.25}$ & $0.9$ & $275\text{--}412$ \\
NGC\,0628 & $0.77^{+0.01}_{-0.01}$ & $-0.72^{+0.10}_{-0.09}$ & $1.67^{+0.19}_{-0.18}$ & $0.57^{+0.02}_{-0.03}$ & $0.75^{+0.01}_{-0.01}$ & $0.77^{+0.04}_{-0.04}$ & $2.1$ & $114\text{--}210$ \\
NGC\,1087 & $0.73^{+0.01}_{-0.01}$ & $-0.41^{+0.14}_{-0.13}$ & $2.34^{+0.22}_{-0.22}$ & $0.72^{+0.03}_{-0.03}$ & $0.69^{+0.01}_{-0.01}$ & $1.03^{+0.05}_{-0.05}$ & $0.9$ & $254\text{--}530$ \\
NGC\,1385 & $0.73^{+0.01}_{-0.01}$ & $-0.75^{+0.10}_{-0.11}$ & $1.78^{+0.16}_{-0.19}$ & $0.56^{+0.02}_{-0.02}$ & $0.78^{+0.02}_{-0.01}$ & $0.72^{+0.04}_{-0.04}$ & $3.0$ & $234\text{--}501$ \\
NGC\,1566 & $0.67^{+0.00}_{-0.00}$ & $-0.62^{+0.01}_{-0.01}$ & $2.11^{+0.02}_{-0.02}$ & $0.59^{+0.00}_{-0.00}$ & $0.76^{+0.00}_{-0.00}$ & $0.77^{+0.01}_{-0.01}$ & $2.1\text{--}3.0$ & $232\text{--}2084$ \\
NGC\,1792 & $0.67^{+0.01}_{-0.01}$ & $-0.90^{+0.05}_{-0.05}$ & $1.70^{+0.07}_{-0.08}$ & $0.64^{+0.02}_{-0.02}$ & $0.82^{+0.01}_{-0.01}$ & $0.78^{+0.03}_{-0.03}$ & $6.1$ & $314\text{--}707$ \\
NGC\,2090 & $0.75^{+0.01}_{-0.01}$ & $-0.39^{+0.19}_{-0.19}$ & $2.32^{+0.33}_{-0.32}$ & $0.63^{+0.05}_{-0.05}$ & $0.69^{+0.02}_{-0.02}$ & $0.91^{+0.08}_{-0.08}$ & $0.9$ & $154\text{--}256$ \\
NGC\,2283 & $0.73^{+0.02}_{-0.01}$ & $-0.95^{+0.25}_{-0.25}$ & $1.44^{+0.41}_{-0.47}$ & $0.79^{+0.05}_{-0.05}$ & $0.76^{+0.02}_{-0.02}$ & $1.04^{+0.10}_{-0.10}$ & $2.1$ & $179\text{--}338$ \\
NGC\,2903 & $0.65^{+0.01}_{-0.01}$ & $-0.82^{+0.02}_{-0.02}$ & $1.87^{+0.03}_{-0.04}$ & $0.73^{+0.01}_{-0.01}$ & $0.76^{+0.00}_{-0.00}$ & $0.95^{+0.01}_{-0.01}$ & $2.1\text{--}4.6$ & $145\text{--}1018$ \\
NGC\,2997 & $0.77^{+0.00}_{-0.01}$ & $-0.71^{+0.05}_{-0.05}$ & $1.71^{+0.10}_{-0.11}$ & $0.61^{+0.01}_{-0.01}$ & $0.73^{+0.00}_{-0.00}$ & $0.82^{+0.01}_{-0.01}$ & $2.1$ & $245\text{--}777$ \\
NGC\,3059 & $0.74^{+0.01}_{-0.01}$ & $-0.64^{+0.10}_{-0.11}$ & $1.93^{+0.19}_{-0.18}$ & $0.52^{+0.02}_{-0.02}$ & $0.76^{+0.02}_{-0.02}$ & $0.67^{+0.04}_{-0.04}$ & $2.1$ & $265\text{--}588$ \\
NGC\,3507 & $0.67^{+0.02}_{-0.02}$ & $0.03^{+0.36}_{-0.42}$ & $3.04^{+0.51}_{-0.60}$ & $0.53^{+0.16}_{-0.17}$ & $0.68^{+0.09}_{-0.07}$ & $0.67^{+0.29}_{-0.31}$ & $0.9$ & $343\text{--}514$ \\
NGC\,3521 & $0.74^{+0.00}_{-0.00}$ & $-0.85^{+0.03}_{-0.03}$ & $1.55^{+0.07}_{-0.06}$ & $0.56^{+0.01}_{-0.01}$ & $0.76^{+0.00}_{-0.00}$ & $0.74^{+0.01}_{-0.01}$ & $2.1\text{--}4.6$ & $173\text{--}481$ \\
NGC\,3621 & $0.72^{+0.00}_{-0.00}$ & $-0.61^{+0.03}_{-0.03}$ & $2.04^{+0.04}_{-0.05}$ & $0.42^{+0.01}_{-0.01}$ & $0.79^{+0.01}_{-0.01}$ & $0.53^{+0.01}_{-0.01}$ & $3.0\text{--}6.1$ & $133\text{--}513$ \\
NGC\,3627 & $0.70^{+0.01}_{-0.00}$ & $-1.14^{+0.02}_{-0.02}$ & $1.26^{+0.05}_{-0.05}$ & $0.73^{+0.00}_{-0.00}$ & $0.79^{+0.00}_{-0.00}$ & $0.90^{+0.01}_{-0.01}$ & $\geq2.1$ & $198\text{--}2107$ \\
NGC\,4254 & $0.73^{+0.01}_{-0.01}$ & $-0.59^{+0.05}_{-0.05}$ & $2.03^{+0.08}_{-0.08}$ & $0.52^{+0.01}_{-0.01}$ & $0.75^{+0.01}_{-0.01}$ & $0.69^{+0.02}_{-0.02}$ & $2.1\text{--}3.0$ & $254\text{--}699$ \\
NGC\,4298 & $0.75^{+0.01}_{-0.01}$ & $-0.58^{+0.07}_{-0.06}$ & $2.00^{+0.13}_{-0.13}$ & $0.53^{+0.01}_{-0.01}$ & $0.74^{+0.01}_{-0.01}$ & $0.71^{+0.02}_{-0.02}$ & $2.1$ & $253\text{--}651$ \\
NGC\,4303 & $0.74^{+0.01}_{-0.01}$ & $-0.48^{+0.04}_{-0.05}$ & $2.19^{+0.08}_{-0.10}$ & $0.54^{+0.01}_{-0.01}$ & $0.72^{+0.00}_{-0.00}$ & $0.75^{+0.01}_{-0.01}$ & $1.4\text{--}2.1$ & $329\text{--}1647$ \\
NGC\,4321 & $0.70^{+0.01}_{-0.01}$ & $-1.14^{+0.08}_{-0.08}$ & $1.26^{+0.12}_{-0.13}$ & $0.65^{+0.02}_{-0.01}$ & $0.79^{+0.01}_{-0.01}$ & $0.82^{+0.03}_{-0.03}$ & $\geq2.1$ & $265\text{--}590$ \\
NGC\,4536 & $0.61^{+0.01}_{-0.01}$ & $-0.80^{+0.05}_{-0.05}$ & $1.97^{+0.07}_{-0.06}$ & $0.68^{+0.02}_{-0.02}$ & $0.77^{+0.01}_{-0.01}$ & $0.87^{+0.03}_{-0.03}$ & $\geq2.1$ & $236\text{--}827$ \\
NGC\,4540 & $0.72^{+0.02}_{-0.02}$ & $-0.13^{+0.35}_{-0.34}$ & $2.79^{+0.56}_{-0.56}$ & $0.64^{+0.12}_{-0.12}$ & $0.71^{+0.04}_{-0.04}$ & $0.87^{+0.18}_{-0.19}$ & $0.9$ & $229\text{--}390$ \\
NGC\,4579 & $0.74^{+0.01}_{-0.01}$ & $-0.88^{+0.10}_{-0.09}$ & $1.52^{+0.18}_{-0.20}$ & $0.57^{+0.02}_{-0.01}$ & $0.78^{+0.01}_{-0.01}$ & $0.73^{+0.03}_{-0.03}$ & $3.0$ & $407\text{--}1222$ \\
NGC\,4654 & $0.78^{+0.01}_{-0.01}$ & $-0.45^{+0.13}_{-0.14}$ & $2.16^{+0.25}_{-0.29}$ & $0.76^{+0.02}_{-0.01}$ & $0.72^{+0.01}_{-0.01}$ & $1.06^{+0.03}_{-0.03}$ & $1.4$ & $373\text{--}852$ \\
NGC\,4689 & $0.77^{+0.01}_{-0.01}$ & $-0.56^{+0.09}_{-0.08}$ & $1.96^{+0.17}_{-0.16}$ & $0.55^{+0.02}_{-0.02}$ & $0.74^{+0.01}_{-0.01}$ & $0.75^{+0.03}_{-0.03}$ & $2.1$ & $175\text{--}393$ \\
NGC\,4781 & $0.73^{+0.01}_{-0.01}$ & $-0.45^{+0.07}_{-0.07}$ & $2.28^{+0.12}_{-0.13}$ & $0.63^{+0.01}_{-0.01}$ & $0.70^{+0.01}_{-0.01}$ & $0.90^{+0.02}_{-0.02}$ & $1.4$ & $148\text{--}428$ \\
NGC\,4826 & $0.76^{+0.01}_{-0.01}$ & $-0.76^{+0.08}_{-0.08}$ & $1.66^{+0.16}_{-0.18}$ & $0.61^{+0.01}_{-0.01}$ & $0.74^{+0.01}_{-0.01}$ & $0.82^{+0.02}_{-0.02}$ & $2.1$ & $58\text{--}231$ \\
NGC\,4941 & $0.74^{+0.01}_{-0.01}$ & $-0.88^{+0.03}_{-0.03}$ & $1.53^{+0.06}_{-0.07}$ & $0.42^{+0.01}_{-0.01}$ & $0.81^{+0.01}_{-0.01}$ & $0.52^{+0.01}_{-0.01}$ & $\geq3.0$ & $255\text{--}2982$ \\
NGC\,5248 & $0.70^{+0.01}_{-0.01}$ & $-0.61^{+0.04}_{-0.05}$ & $2.07^{+0.07}_{-0.08}$ & $0.84^{+0.01}_{-0.01}$ & $0.72^{+0.00}_{-0.00}$ & $1.16^{+0.02}_{-0.02}$ & $1.4\text{--}2.1$ & $195\text{--}606$ \\
NGC\,6300 & $0.64^{+0.01}_{-0.01}$ & $-0.79^{+0.05}_{-0.05}$ & $1.93^{+0.07}_{-0.07}$ & $0.60^{+0.01}_{-0.01}$ & $0.78^{+0.01}_{-0.01}$ & $0.76^{+0.03}_{-0.03}$ & $\geq2.1$ & $135\text{--}472$ \\
NGC\,7496 & $0.68^{+0.01}_{-0.01}$ & $-1.06^{+0.13}_{-0.13}$ & $1.45^{+0.20}_{-0.21}$ & $0.85^{+0.03}_{-0.03}$ & $0.75^{+0.01}_{-0.01}$ & $1.12^{+0.06}_{-0.05}$ & $1.4\text{--}4.6$ & $318\text{--}635$ \\
\midrule
\midrule
{\scriptsize\textit{Median}} & {\scriptsize $0.73^{+0.03}_{-0.06}$} & {\scriptsize $-0.71^{+0.27}_{-0.18}$} & {\scriptsize $1.93^{+0.35}_{-0.40}$} & {\scriptsize $0.61^{+0.13}_{-0.07}$} & {\scriptsize $0.75^{+0.04}_{-0.04}$} & {\scriptsize $0.78^{+0.25}_{-0.07}$} & {\scriptsize $-$} & {\scriptsize $234\text{--}590$} \\
\bottomrule
\caption{This table lists the 31 selected galaxies, which are the galaxies with a reliable kinematic fit and a smooth analytic fit to the rotation curve in Lang et al. (2020) and an inclination $i + \sigma_i \leq 70^\circ$. It contains the cascade ratio $\beta_{\mathrm{SL}}$, singularity exponent $\alpha$, Hausdorff dimension $D$, and the scaling exponents $\zeta_2$ and $\zeta_3$, and ESS exponents $\beta_2$ from the spatial bootstrap, along with their errors as the 16th and 84th percentiles of the bootstrap distribution. The row labelled \textit{Median} gives the median across the galaxy population of each of the exponents, and the corresponding errors report the distance from that median to the 16th and 84th percentiles. The $\mathcal{M}_s$ classification column gives each galaxy's classification against the six Pan and Scannapieco (2011) Mach number curves. A single Mach number is provided where one curve fits well, else the range it falls between is reported. Lastly, the table includes the range in which the scaling exponents are fitted for each galaxy.}
\label{tab:scaling-exponents}
\end{longtable*}
\endgroup
\clearpage

\subsection{Linear interdependence in molecular clouds}
\begin{figure}[h]
    \centering
    \includegraphics[width=0.9\linewidth]{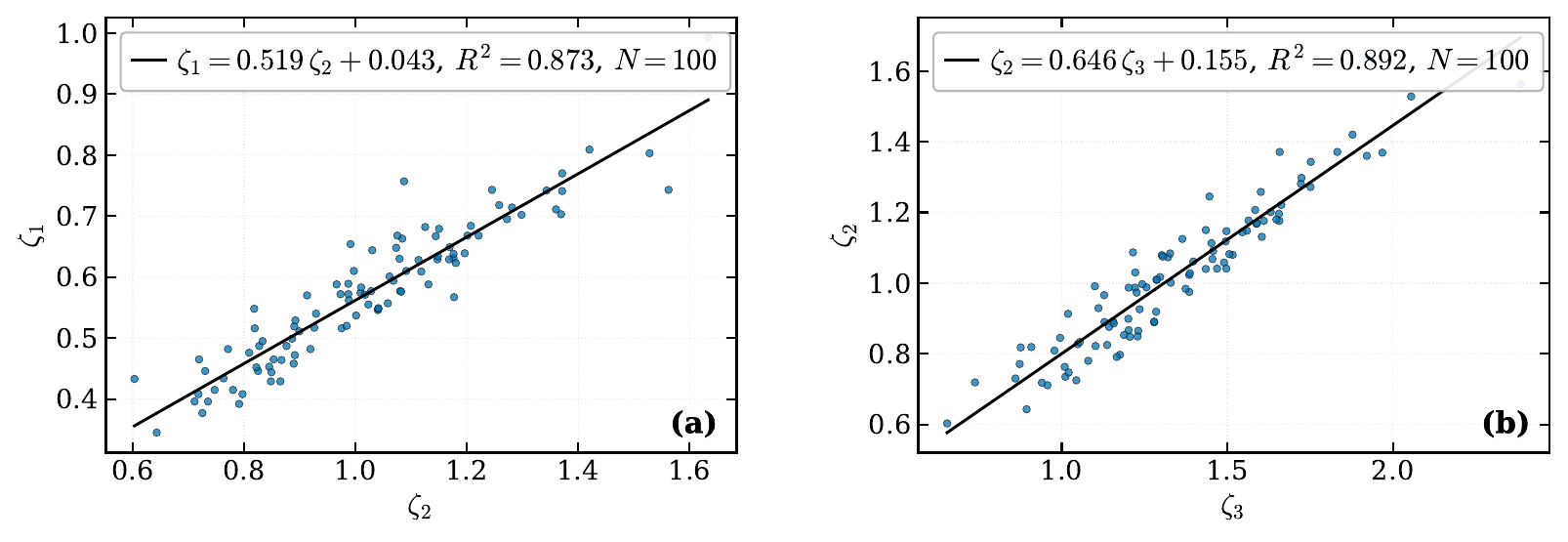}
    \caption{The first to third order scaling exponents $\zeta_p$ for 100 molecular clouds computed by \citet{ma_examining_2025} plotted against their consecutive exponent $\zeta_{p+1}$. An ordinary linear least-squares fit provides slopes of 0.519 and 0.646, which are close to the $p/(p+1)$ reference coefficients.}
    \label{fig:GMCinterd}
\end{figure}

In this appendix we perform the same population-level analysis, as in the letter for PHANGS-ALMA (also explained in the previous section of this appendix), in 100 molecular clouds studied in \citet{ma_examining_2025}.  \autoref{fig:GMCinterd} shows the scaling exponents for the 100 molecular clouds that showed a proper power law, computed by \citet{ma_examining_2025}. Because the bootstrap covariance of these scaling exponents is not readily available, the consecutive orders are fitted with ordinary linear least squares providing slopes of 0.519 and 0.646, which are close to the $p/(p+1)$ reference coefficients. The plotted cloud estimates give $\Delta_1=\zeta_1-\zeta_2/2>0$ in 96 of 100 clouds and $\Delta_2=\zeta_2-2\zeta_3/3>0$ in 99 of 100, with median $\zeta_2/\zeta_3\simeq0.749$. These calculations use the exponent coordinates retained in the original figure in \citet{ma_examining_2025} and shared second-order values. The sample comprises the 100 of 167 clouds with accepted power-law fits.  The relative exponents of these clouds mostly lie close to the Kolmogorov-Burgers prediction for supersonic turbulence \cite{ma_examining_2025,boldyrevKolmogorovBurgersModelStarforming2002}, consistent with a shock-sheet interpretation in that fixed model. This provides a qualitative counterpart to the PHANGS comparison at larger scales. Because \citet{ma_examining_2025} report only orders one to three, we do not assign its 100 clouds the six-order simulation based Mach estimates used for PHANGS, or infer a separate generalized $D$ for each cloud.

\end{document}